\documentclass[iop,apj]{emulateapj}         

\usepackage{graphics}
\usepackage[usenames,dvipsnames]{color}
\usepackage[normalem]{ulem}  

\newcommand{\um}{${\rm \mu m}$~}  
\newcommand{\mm}{${\rm \mu m}$}

\long\def\symbolfootnote[#1]#2{\begingroup% 
\def\thefootnote{\fnsymbol{footnote}}\footnote[#1]{#2}\endgroup}

\slugcomment{} 
\shortauthors{Su et al.}

\begin{document}

\title{The Inner Debris Structure in the Fomalhaut$^{\ast}$ Planetary System}

\author{Kate Y. L. Su\altaffilmark{1,6}, 
  George H. Rieke\altaffilmark{1}, 
  Denis Defr\`ere\altaffilmark{1},
  Kuo-Song Wang\altaffilmark{2},
  Shih-Ping Lai\altaffilmark{3,6},
  David J. Wilner\altaffilmark{4}, 
  Rik van Lieshout\altaffilmark{5}, and 
  Chin-Fei Lee\altaffilmark{2}
  }   

\affil{$^1$Steward Observatory, University of Arizona, Tucson, AZ 85721, USA; ksu@as.arizona.edu} 

\affil{$^2$Institute of Astronomy and Astrophysics, Academia Sinica, P.O. Box 23-141, Taipei 106, Taiwan}

\affil{$^3$Institute of Astronomy, National Tsing Hua University (NTHU), Hsinchu 30013, Taiwan}

\affil{$^4$Harvard-Smithsonian Center for Astrophysics 60 Garden Street, Cambridge, MA 02138, USA}

\affil{$^5$Anton Pannekoek Institute for Astronomy, University of Amsterdam, Science Park 904, 1098 XH Amsterdam, The Netherlands}

\altaffiltext{$^{\ast}$}{Fomalhaut is a triple system; here we refer the Fomalhaut planetary system as the one around
the primary star Fomalhaut A.}
 \altaffiltext{6}{Visiting Scholar, Institute of Astronomy and Astrophysics, Academia Sinica, P.O. Box 23-141, Taipei 106, Taiwan}

\journalinfo{accepted for publication in ApJ}
\submitted{} 

\begin{abstract}

Fomalhaut plays an important role in the study of debris disks and
small bodies in other planetary systems. The proximity and luminosity
of the star make key features of its debris, like the water ice-line,
accessible. Here we present ALMA cycle 1, 870 \um (345 GHz)
observations targeted at the inner part of the Fomalhaut system with a
synthesized beam of 0\farcs45$\times$0\farcs37 ($\sim$3 AU linear
resolution at the distance of Fomalhaut) and a rms of 26 $\mu$Jy
beam$^{-1}$. The high angular resolution and sensitivity of the ALMA
data enable us to place strong constraints on the nature of the warm
excess revealed by {\it Spitzer} and {\it Herschel} observations. We
detect a point source at the star position with a total flux
consistent with thermal emission from the stellar photosphere. No
structures that are brighter than 3$\sigma$ are detected in the
central 15 AU$\times$15 AU region. Modeling the spectral energy
distribution using parameters expected for a dust-producing
planetesimal belt indicates a radial location in the range $\sim$8--15
AU. This is consistent with the location where ice sublimates in
Fomalhaut, i.e., an asteroid-belt analog.  The 3$\sigma$ upper limit
for such a belt is $<$1.3 mJy at 870 $\mu$m. We also interpret the 2
and 8--13 \um interferometric measurements to reveal the structure in
the inner 10 AU region as dust naturally connected to this proposed
asteroid belt by Poynting-Robertson drag, dust sublimation, and
magnetically trapped nano grains.

\end{abstract} 


\section{Introduction}

The Fomalhaut debris disk has become the prototype for understanding
the complexity of disk structures and their relationship to planetary
systems.  Due to its proximity (7.7 pc) and luminosity (17.4
L$_{\sun}$), many key features of its debris are easily accessible
through optical, infrared and millimeter ground- and space-based
facilities. The prominent planetesimal ring located at $\sim$140 AU,
analogous to the Kuiper belt in our Solar System, is particularly
well-studied \citep{holland98,kalas05,acke12,ricci12}. The sharp
boundaries of this cold ($\sim$50 K) ring suggest gravitational
shaping by unseen planets \citep{quillen06,chiang09, boley12}. Intense
collisions among the planetesimals in this ring grind large (parent)
bodies into fine dust, forming a disk halo composed of $\sim$$\mu$m
size grains outside the cold ring \citep{acke12,kalas13}. The
Fomalhaut disk has a warm ($\sim$150 K), unresolved component near the
star, first indicated by {\it Spitzer} 24 $\mu$m imaging
\citep{stapelfeldt04} and later confirmed by {\it Herschel} imaging
\citep{acke12}. This warm excess has been suggested as an ice-line
planetesimal belt, analogous to the asteroid belt in our Solar System
\citep{su13}. Furthermore, the Fomalhaut system is thought to have a
very hot ($\sim$1500 K) exozodiacal dust population at $\sim$0.1 AU,
revealed by ground-based interferometry at 2 $\mu$m ($K$-band)
\citep{absil09}. Interferometric data at $\sim$10 \um detect a hot
component near $\sim$1--2 AU
\citep{mennesson13,lebreton13}. Therefore, the Fomalhaut system is the
only system known to possess all five debris zones (very hot, hot,
warm, cold, and halo dust) defined by \citet{su14}.

In addition to the debris system around Fomalhaut, a combination of
imaging where the warm component can be spatially separated from the
cold component and mid-infrared spectroscopy has revealed similar
inner warm components in other nearby systems, for example, $\epsilon$
Eri: \citep{backman09,greaves14}, HR 8799: \citep{su09,matthews14},
Vega: \citep{su13}. $\sim$20\% of unresolved debris disks show such
warm components in their disk spectral energy distributions (SEDs)
\citep{morales11,ballering13,chen14,kennedy14}. Leftover parent bodies
near the star generally have higher collisional velocities, i.e.,
shorter dynamical time scales, compared to those of more distant ones,
as suggested in the fading of the warm components after a few hundred
Myr (e.g., \citealt{gaspar13}). These components may be analogous to
our asteroid belt but with far greater mass; however, compared to the
cold Kuiper-belt analogs they are much harder to characterize in
exoplanetary systems. Identification and characterization of such warm
components require very accurate extrapolation of the stellar spectrum
and precise subtraction.

Not surprisingly, although it can be considered the prototype, the
nature of the Fomalhaut warm component is
controversial. \citet{acke12} propose that it originates from
free-free emission of a postulated stellar wind that also creates the
hot excess revealed in near-infrared interferometry. However,
\citet{su13} draw from all the resolved images and the spectroscopy to
argue that it is a planetesimal ring placed near the ice line and
presumably created by processes occurring in other debris systems as
well. In that case the Fomalhaut debris disk resembles the debris
structures in our Solar System -- a dense Kuiper belt whose inner edge
is maintained by massive planets \citep{liou99} and a more tenuous
asteroid belt containing dust structure (zodiacal cloud) determined by
both the giant and terrestrial planets \citep{dermott94}.

There is a tentative detection of an excess in the ALMA cycle 0 image
of Fomalhaut \citep{boley12}, but less than  the prediction by the
free-free model, and the star was at the edge of the primary beam so
the excess flux is subject to the large uncertainty of the primary
beam correction. Here we present a high angular resolution and
sensitivity ALMA cycle 1 image centered at the star position, and
provide a strong constraint on the amount of excess emission in the
inner part of the Fomalhaut system.

Section 2 of this paper describes the details of the ALMA observation
and data reduction. In Section 3, we discuss the 870 \um measurement
near the star, evaluate the expected photospheric value of Fomalhaut
at this wavelength, and rule out any compact free-free emission
originated from the star as the possible source of excess emission
detected by {\it Spitzer} and {\it Herschel}. In Section 4.1 we review
the properties of the inner 20 AU region in Fomalhaut by assessing all
available excess measurements and discussing their uncertainties. We
conclude that a warm excess at the ice sublimation temperature
($\sim$150 K, i.e., an asteroid-belt analog) is robustly detected,
independent of the uncertainty of photospheric subtraction and the
presence of a tentative, hot ($\sim$500 K) excess. In Section 4.2, we
present SED models for the ice-line asteroid belt around Fomalhaut and
set constraints on its total flux at 870 $\mu$m, and we also show that
a 13-AU narrow belt might be detected at $\lesssim$2$\sigma$ levels in
the cycle 1 ALMA data. In Section 4.3, we propose a new interpretation
for the hot excess detected by the Keck Interferometric Nuller (KIN)
via linking the proposed asteroid belt to the dust structure in the
inner 10 AU region. Additional validation in terms of the derived dust
mass and the sensitivity of the adapted grain parameters to our model
are discussed in Section 5, followed by our conclusions in Section 6.

\begin{deluxetable*}{llccccc}
\tablewidth{0pc}
\tablecaption{ALMA Cycle 1 Observations of Fomalhaut\label{tbl:obs}}
\tablehead{ 
\colhead{EB } & \colhead{UT Date} & \multicolumn{3}{c}{Calibrators } & \colhead{\# of} & \colhead{PWV}\\
\colhead{code}        & \colhead{}    & \colhead{Passband} & \colhead{Flux} & \colhead{Phase} & \colhead{anntena} &{[mm]}
}
\startdata 
X4a7	 & 2014 Apr 29	 & J2258$-$2758	 & J2258$-$279	 & J2250$-$2806	 & 35   & 1.43 \\
X11ad	 & 2014 May 21	 & J2258$-$2758	 & J2258$-$279	 & J2250$-$2806	 & 32   & 1.10 \\
X135f	 & 2014 May 21	 & J2258$-$2758	 & J2258$-$279	 & J2250$-$2806	 & 32   & 1.24 \\	
Xe46	 & 2014 May 26	 & J2056$-$4714	 & J2258$-$279	 & J2258$-$2758	 & 25   & 0.67 \\
X119f	 & 2014 May 26	 & J2258$-$2758	 & J2258$-$279	 & J2250$-$2806	 & 25   & 0.72 \\      
\enddata 
\end{deluxetable*}

\section{Observations and Data Reduction} 
\label{obs}

Fomalhaut was observed with the ALMA 12-m array in Cycle 1 (program
2012.1.00238.S, PI: Lai) in five execution blocks at band 7. Table 1
summarizes the detailed information for each execution including the
bandpass, amplitude, phase calibrators, and the estimated precipitable
water vapour (PWV). The data taken on the second execution of 2014 May
21 did not pass the quality assurance procedure, thus are not included
in the imaging and analysis study. A total on-source time of 172
minutes was achieved. The phase center is located at the star position
(RA=22:57:39.409, Dec=$-$29:37:22.423) corrected for proper motion at
the epoch of the observation. Four 2-GHz wide spectral windows
centered at 337.987 GHz, 339.924 GHz, 350.002 GHz and 352.002 GHz were
used to measure the continuum emission with dual linear polarizations
from the target source. The primary beam size (the half-power width of
the Gaussian beam) is 18\arcsec\ at 345 GHz.  The overall
$uv$-coverage ranges from 23.0 k$\lambda$ to 737.7 k$\lambda$. Thus,
our data are not sensitive to smooth emission structures larger than
$\sim$9\arcsec.  The width of the outer belt is estimated to be
$\sim$13--19 AU \citep{boley12}, which is mostly resolved out by our
high resolution ($\sim$3 AU in linear scale); therefore, our data only
show a very faint trace of the outer belt along the minor-axis
direction.  The quasar J2258$-$279 was used as the amplitude
calibrator in all execution blocks. The flux uncertainty is estimated
to be 10\%.

All the data reduction and imaging processes were carried out with
CASA 4.2.1 \citep{mcmullin07}. The final continuum image was obtained
by performing multi-frequency-synthesis (mfs) imaging. Clean boxes of
the expected cold belt location (generated with reference to the {\it
HST} image) and the central 5\arcsec\ for the warm belt, were used in
the image deconvolution process.  A final image with effective
bandwidth of 8 GHz centered at 345 GHz was generated. The synthesized
beam is 0\farcs45$\times$0\farcs37 with position angle (P.A.) of
82.3\arcdeg. The rms noise in the final mfs image is 26 $\mu$Jy
beam$^{-1}$.

\section{ALMA Observational Results}

\subsection{870 \um Measurement}

\begin{figure*} 
  \figurenum{1}
  \label{fom_alma_cycle1} 
 \epsscale{1.15} 
  \plotone{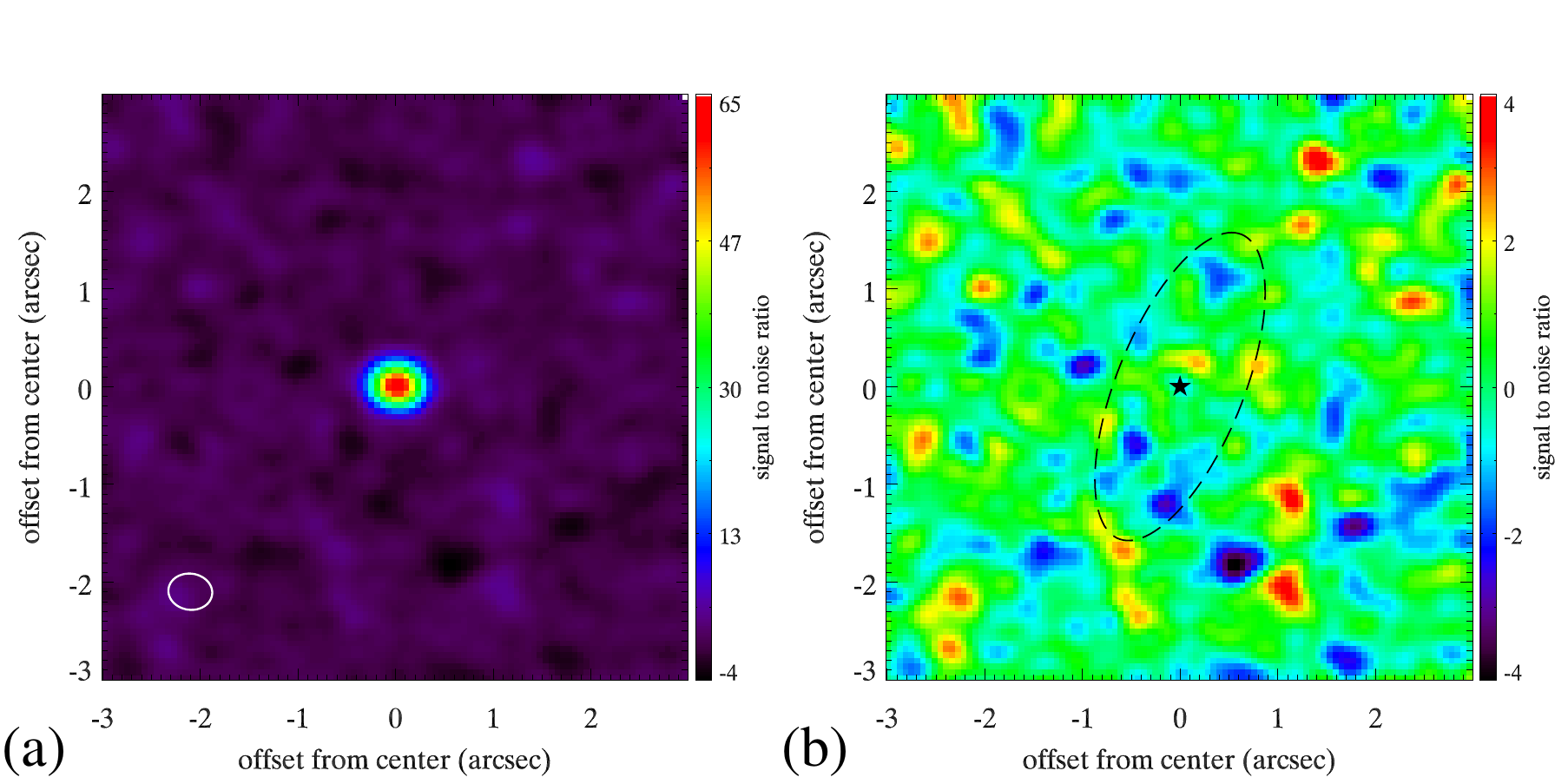}
  \caption{(a) Cycle 1 ALMA map of the central
6\arcsec$\times$6\arcsec\ region centered at the stellar position in
the Fomalhaut system. The synthesized beam, shown as a white ellipse,
has a FWHM of 0\farcs45$\times$0\farcs37, $\sim$3 AU linear resolution
at the distance of Fomalhaut (7.7 pc). Our cycle 1 observations reach
a rms of 26 $\mu$Jy~beam$^{-1}$. (b) Star-subtracted residual map. No
structures that are brighter than 3$\sigma$ are detected in the
2\arcsec$\times$2\arcsec\ (15 AU$\times$15 AU) central region. The
expected disk circumference of a 13-AU narrow belt, which has the same
inclination and position angles as the outer cold belt but centered at
the star, is marked as the dashed ellipse. A few $\sim$2$\sigma$ blobs
are found along the ellipse.  }
\end{figure*}

Figure \ref{fom_alma_cycle1}(a) shows the central
6\arcsec$\times$6\arcsec\ region of the map. A bright source, detected
at the star position after taking the star's proper motion into
account, has a full-width-half-maximum (FWHM) of
0\farcs46$\times$0\farcs38 at P.A. of 86.4\arcdeg, consistent with
being the star (also see Sec 3.2). To uncover any faint structure that
might be present around the star, we fitted a single point source
model to the data and subtracted it from the original data in the $uv$
plane, then generated the residual map. We used {\it uvmodelfit\ } in
CASA to determine the best-fit parameters for the point source: a
total flux of 1.789$\pm$0.037 mJy. The quoted uncertainty is based on
the fit, i.e., this number does not include other errors from flux
calibration. The residual map is shown Figure
\ref{fom_alma_cycle1}(b). No structures that are brighter than
3$\sigma$ are detected in the 2\arcsec$\times$2\arcsec\ (15
AU$\times$15 AU) region around the star after subtracting the best-fit
point source model. Assuming the inner belt has a similar orientation
as the outer belt, one would expect the disk ansae to be along the
P.A. of 156\arcdeg. There are two $\sim$2$\sigma$ blobs along that
angle: one is $\sim$0\farcs4 (3 AU) north-west of the star, and the
other is $\sim$1\farcs8 (14 AU) south-east of the star. Based on our
SED model (see Section 4.2), the likely location of the inner belt is
at radii of $\sim$8--15 AU, suggesting the south-east blob could be
part of the inner belt. Assuming the inner belt is also inclined by
66\arcdeg\ like the outer one, but centered at the star (no offset),
the ellipse on Figure \ref{fom_alma_cycle1}(b) marks the boundary of a
putative 13-AU inner belt. Interestingly, there are a few 1--2$\sigma$
blobs along the expected disk circumference, suggesting this 13-AU
inner belt might be detected at 1-2$\sigma$ levels (more discussion is
given in Section 4.2). Although the 1-2$\sigma$ positive blobs along
the non-offset, expected 13-AU disk circumference are intriguing, we
note that there are also many $-$2$\sigma$ blobs in the data as well.
In summary, we do not find any significant extended structure around
the star, and the reality of the putative 13-AU belt needs future
confirmation.

\subsection{Expected Photosphere Brightness at 870 \um and A Constraint on the Unresolved Excess}

\citet{su13} evaluated all available optical to near-infrared
photometry and determined the stellar photospheric output to be 2.96
Jy in the MIPS 24 \um band. A simple Rayleigh-Jeans extrapolation then
yields an estimate of 2.2 mJy at 870 $\mu$m, significantly brighter
(by $\sim$23\%) than the point source measured at the star
position. The discrepancy likely comes from two sources: (1) the
uncertainty of the ALMA absolute calibration, and (2) the uncertainty
of Rayleigh-Jeans extrapolation.

The accuracy in absolute calibration for ALMA Cycle 2 is set at 10\%
for Band 7 (J. Mangum, priv. communication). To verify the
calibration, we used the solar analog method \citep{johnson65,rieke08}
by comparing measurements of $\alpha$ Cen A with those of the Sun. We
did this in two parts. First, we verified $\alpha$ Cen A is a solar
analog by comparing its colors with those of the Sun. Then, we
computed the submillimeter flux of $\alpha$ Cen A using the solar
relation between measured 24 \um and submillimeter. We derived colors
of $V-K_s$=1.531$\pm$0.023, $V-[24]$=1.584$\pm$0.027, and
$K_s-[24]$=0.053$\pm$0.032 using the available $K$-band measurement
from \citet{engels81}, $V$-band from \citet{alekseeva96},
\citet{mermilliod91} and {\it Hipparcos}, and 24 \um measurement of
30.84 Jy from \citet{wiegert14} after proper photometric system
transformation and calibration to the zero points defined by
\citet{rieke08}. These colors are virtually identical to those of the
Sun and establish $\alpha$ Cen A as a valid solar analog star for the
purpose of calibration.

We then used measurements of the solar brightness temperature to
connect the mid-infrared measurements to the ALMA calibration at 870
$\mu$m. We adopt the brightness temperature at 24 $\mu$m to be 4625 K
(Model M in \citealt{vernazza76}), and the brightness temperature at
870 \um to be 5470 K \citep{loukitcheva00}. Taking 30.84$\pm$0.76 Jy
as the 24 \um flux of $\alpha$ Cen A \citep{wiegert14}, the scaling
from the solar brightness temperatures suggests $\alpha$ Cen A has a
flux density of 28.9 mJy at 872 $\mu$m. The error associated with this
number is not straightforward to  determine, but can be estimated as 
the quadratically combined error of 7\% ($\pm$300 K (6\%) in the solar
870 \um brightness temperature, $\sim$2\% in the solar 24 \um
brightness temperature, and 2.5\% of the $\alpha$ Cen A 24 \um
measurement). The measurement of 26.1$\pm$0.2 mJy for $\alpha$ Cen A
from \citet{liseau15} then suggests that the solar analog calibration
is brighter than the ALMA measurement by 10.7$\pm$7\%, which confirms
the 10\% ALMA calibration uncertainty estimate is plausible.  
%the ALMA calibration.

If we adjust the measurement of Fomalhaut to this alternative
calibration, we find a flux density of 1.98 mJy (instead of 1.789
mJy), still somewhat fainter (by $\sim$11\%) than the Rayleigh-Jeans
extrapolation from 24 $\mu$m. We conclude that radiative transfer
effects may cause a drop in the brightness temperature of the star,
similar to the behavior of the Sun, which has a brightness temperature
$\ge$5800 K across the visible \citep{vernazza76} but of only 5470 K
at 870 $\mu$m.  The result that Fomalhaut is slightly fainter than the
Rayleigh-Jeans extrapolation value is robust, and this independent
calibration of our measurement also rules out any compact free-free
emission originating from the star as the possible source of excess
emission detected by {\it Spitzer} and {\it Herschel}. 

\section{Analysis}

\subsection{Properties of Inner 20 AU Excess in Fomalhaut} 

In this subsection, we review all the available measurements that can
be used to constrain the excess properties of the inner 20 AU region
in Fomalhaut. These measurements can be categorized in two classes:
(I) resolved measurements where emission from the cold belt is mostly
excluded, and (II) unresolved measurements at the wavelengths where
the cold belt contributes much less emission. In Category (I), the
ground-based interferometric measurements suggest a $K$-band excess of
0.88\%$\pm$0.12\% (\citealt{absil09}, hereafter the VLTI measurement)
and 8--13 \um excess of 0.35\%$\pm$0.10\% (\citealt{mennesson13},
hereafter the KIN measurements) relative to the stellar
photosphere. These interferometric observations can very effectively
isolate the region that is very close to the star ($\lesssim$2 AU in
K-band and $\lesssim$6 AU at 8--13 \mm) from the contribution of the
stellar photosphere.  In addition, the excess emission at 24 and 70
\um inferred from the resolved images of Fomalhaut by {\it Spitzer}
and {\it Herschel} also belongs to the first category. Since the star
and any unresolved excess are both contained within the instrument's
point spread function (PSF), the excess estimate does depend on the
expected photospheric values.

In determining the unresolved excess, an accurate value for the
photospheric flux is essential; it can be determined at $\sim$2\%
levels by fitting the stellar atmospheric models to the optical to
near infrared photometry \citep{su05,engelbracht07}.  At 24 $\mu$m,
an additional unresolved source of 0.6$\pm$0.2 Jy was found in the
early reduction of the MIPS data \citep{stapelfeldt04}. We reduced the
original data with the final in-house reduction pipeline and
calibration, and employed the PSF subtraction method. We find the
unresolved excess at 24 \um is 0.64$\pm$0.13 Jy, consistent with the
old value (but with a smaller uncertainty).  \citet{acke12}
constrained the amount of the unresolved excess by fitting the
resolved images with a SED model using complex dust grains (a mixture
of compositions proposed by \citet{min11}), and derived a total of
0.17$\pm$0.02 Jy for the unresolved excess at 70 $\mu$m. \citet{su13}
employed the PSF subtraction method, and derived a consistent number,
0.17 Jy with upper- and lower-bound fluxes of 0.58 Jy and 0.136 Jy for
the unresolved excess.  The resolution at both {\it Spitzer} 24 \um
and {\it Herschel} 70 \um is very similar ($\sim$6\arcsec), which
places the excess emission mostly from the region within $\sim$20 AU
in projected stellocentric distance.

\setcounter{footnote}{0}

\begin{figure} 
  \figurenum{2}
  \label{fom_innerexcess} 
  \plotone{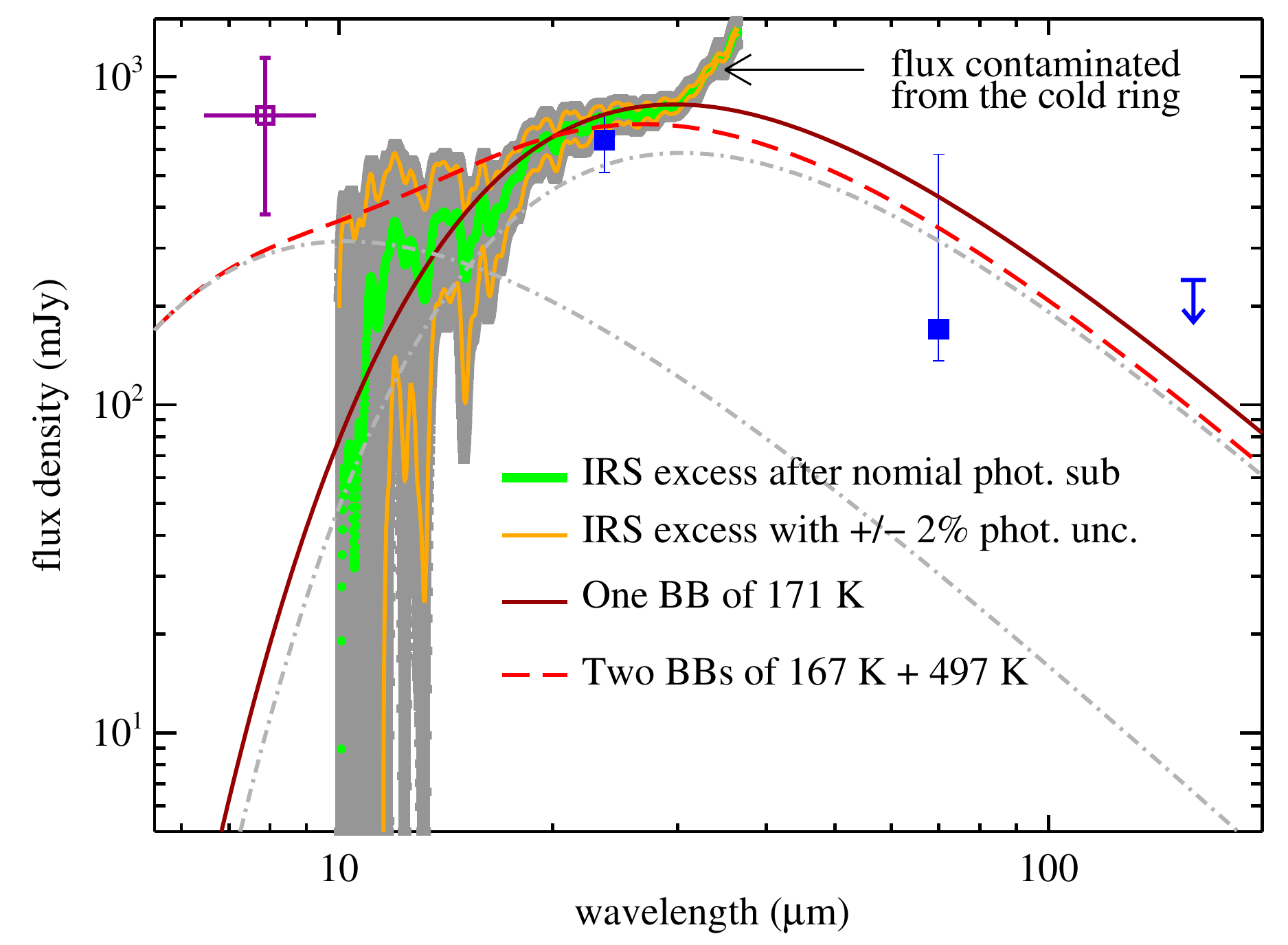}
  \caption{SED of the infrared excess for the inner 20 AU region
around Fomalhaut. The thick green line is from the IRS spectrum
presented in \citet{su13} after the nominal photospheric subtraction.
The gray area represents the uncertainty of the IRS excess, including 2\% of the
photospheric subtraction. The thin orange lines represent the excess IRS spectra
if the photosphere is $\pm$2\% higher/lower than the nominal value.
The filled squares represent the broadband excess emission
estimated from the resolved images, the open square is from
the IRAC measurement (see text for details), and the 3$\sigma$ upper
limit at 160 \um is shown as a downward arrow. With the nominal and 2\% higher
photospheric subtraction, the IRS excess spectra (green and lower 
orange lines) are best described by a blackbody emission of $\sim$170 K 
for wavelengths shorter than 30 $\mu$m. If the photosphere is lower by 2\%,
the resultant IRS excess spectrum is best described by a combination of two 
blackbody temperatures at 167 K and 497 K (two gray dotted dash lines). }
\end{figure}

In Category (II), the {\it Spitzer} IRS spectrum of the Fomalhaut
system presented in \citet{su13} provides good measurements in the
10--30 \um region. The IRS spectrum was extracted and calibrated as a
point source (resolution of $\sim$6\arcsec). With a large slit size of
11\farcs1 in the IRS long-high channel, the flux longward of $\sim$30
\um might be contaminated by the cold ring. Furthermore, the IRS
excess emission is subject to the uncertainty of photospheric
subtraction. To account for this, the uncertainty for the IRS excess
emission includes 2\% of the photospheric emission (added
quadratically) as shown in Figure \ref{fom_innerexcess}. To illustrate
the effect of photospheric subtraction, we also present excess spectra
by artificially scaling our best-determined photospheric model by
$\pm$2\%. The resultant spectra (shown in Figure
\ref{fom_innerexcess}) have higher/lower flux in the 10--15 \um
region, but both are within the estimated uncertainty (gray area). We
augment the IRS excess determination with additional
accurate\footnote{To obtain infrared excesses at the $\sim$2\% level,
using a color difference referenced to a clean, no infrared excess star
(like Sirius) can avoid the uncertainty in photospheric subtraction as
long as both the star of interest and the reference star (Sirius) have
a similar spectral type and accurate photometry. Unfortunately,
ALLWISE and AKARI IRC catalogs do not match the latter criterion
sufficiently well.} broadband photometry. We adopt the accurate PSF
fitting photometry ($\pm$1.5\% level) by \citet{marengo09} for
Fomalhaut and Sirius in the IRAC bands. The color difference between
Fomalhaut and Sirius is 2.36 mag at IRAC 1, 2, 3 bands, but 2.33 mag
at IRAC band 4, suggesting a 3\%$\pm$1.5\% excess at IRAC 8 $\mu$m (a
very broad bandwidth of 36\%).  Since the IRAC measurement was derived
by PSF fitting, the excess most likely originates from the region
within 8 AU (in radius) (resolution of $\sim$2\arcsec\ at IRAC 8
$\mu$m).

One good way to characterize where the bulk of excess emission comes
from is to determine the dominant dust temperature in the excess
spectrum. This dominant dust temperature simply means that the
majority of the dust grains have a similar dust temperature, which is
not subject to a specific grain model even though this temperature is
usually derived by the blackbody formula. Fitting the IRS excess
emission, the inner warm emission is best described by a blackbody
emission of $\sim$170 K (see Figure \ref{fom_innerexcess}); and the
temperature remains the same for the excess spectrum that was
over-subtracted by 2\% more of the photosphere. If the photosphere
level is lower by 2\% compared to the nominal level (i.e.,
under-subtracted), the resultant IRS excess emission needs at
least two temperatures to fit, 167 K and 497 K. Due to the uncertainty
of the photospheric subtraction, we cannot rule out the presence of
$\sim$500 K dust in the system, especially, the two-temperature fit is
a better fit to the IRAC 8 \um point. An alternative fit in this
case is a single blackbody plus a weak (broad) silicate emission
feature (see discussion in Section 4.3.2 and Figure
\ref{complex_model} for some examples).  Even with this tentative
excess emission either at $\sim$500 K or a broad feature, the
10--30 \um excess can only be fitted with a cooler component at 167 K,
suggesting the robustness of the warm excess at the temperature of
water ice sublimation ($\sim$150 K), i.e., an asteroid-belt analog.

\subsection{Parameters for the Asteroid Belt Model}

\begin{figure} \figurenum{3}
  \label{fom_innerexcess_models}
\plotone{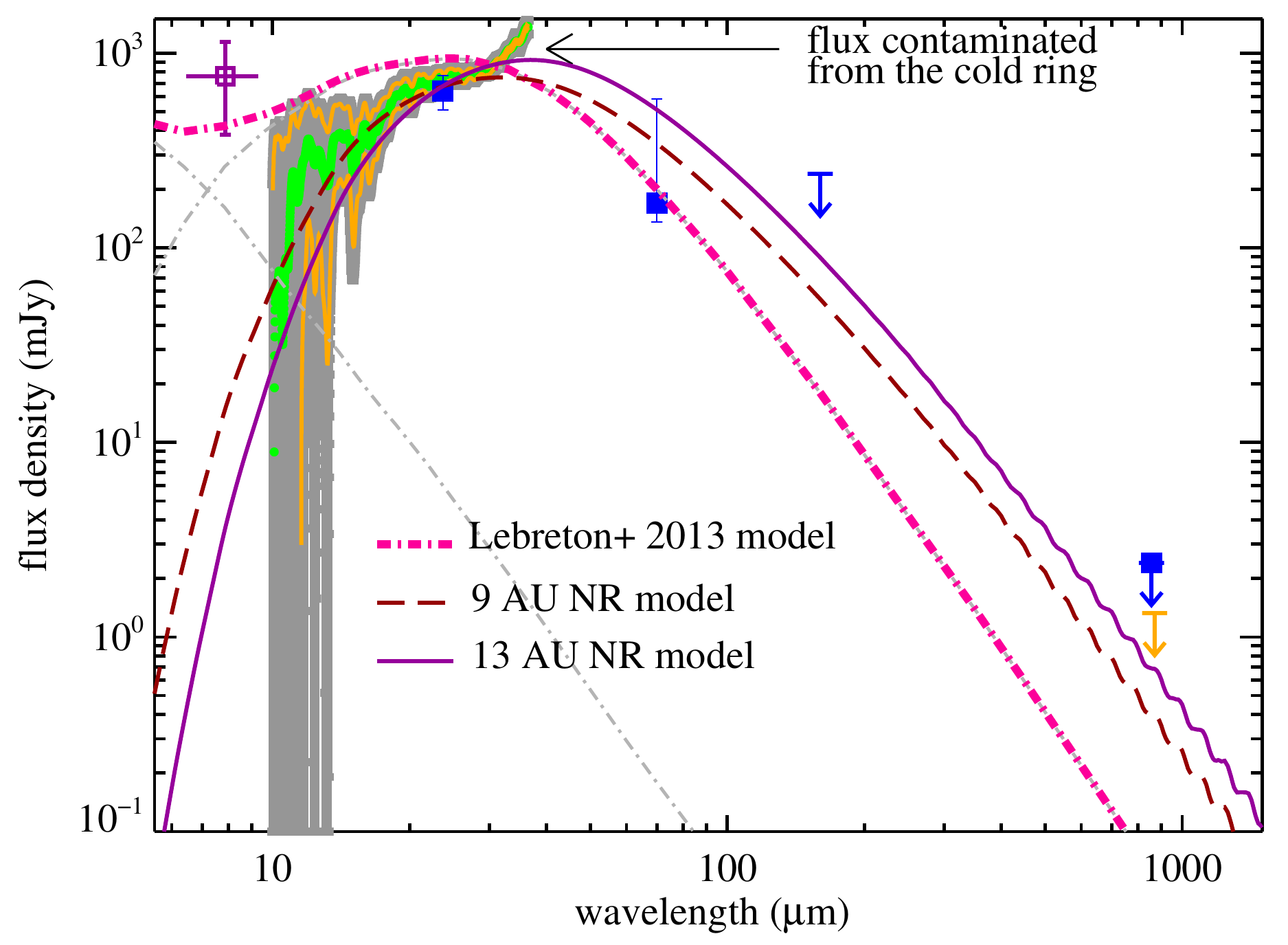}
  \caption{SED of the infrared excess for the inner 20 AU region
around Fomalhaut with various models. The data points are the same as
in Figure \ref{fom_innerexcess}, except for the ALMA Band 7 upper
limits. The measurement from the cycle 0 Band 7 \citep{boley12} is
shown as a filled square with a downward arrow, while the one from
cycle 1 is shown as an orange downward arrow. The model from
\citet{lebreton13} is shown as the thick dot-dash line composed of two
belts (thin dotted-dash lines) at $\sim$0.1 and $\sim$2 AU. Our narrow
ring (NR) models are aimed to fit the excess emission near the ice
line (asteroid-belt analog) and are constrained by data points
longward of 10 $\mu$m.}
\end{figure}

We use a simple optically thin SED model to estimate the physical
location of the asteroid belt. Our SED model is based on the
assumption that the warm ($\sim$170 K) excess originates from a
planetesimal belt located near the water ice line. Therefore, we only
use data longward of $\sim$10 \um to constrain our fits and do not
include the possible hot $\sim$500 K component (see Section 4.3 for
further discussion). In this scenario, dust debris is generated
through collisions of large parent bodies and cascades down to fine
grains (close to the radiation blowout size) with a typical power-law
slope of $-$3.5$\sim$$-$3.7 \citep{wyatt11,gaspar12}. The radiation
blowout size\footnote{The grain size can be parametrized by the $\beta$
value defined as the ratio of the radiation force and the
gravitational force on a particle, i.e., $\beta = \frac{3Q_{\rm pr}
L_{\ast}}{16 \pi c\ G M_{\ast} a \rho_g}$ where $Q_{\rm pr}$ is the
radiation pressure efficiency averaged over the stellar spectrum,
$L_{\ast}$ and $M_{\ast}$ are the stellar luminosity and mass, $G$ is
the gravitational constant, $c$ is light speed, $a$ is the grain
radius, and $\rho_g$ is the grain density. The blowout size ($a_{bl}$)
is defined when $\beta=$ 0.5, i.e., $\frac{a_{bl}}{\mu{\rm m}}=
1.153\frac{Q_{\rm pr}}{1} \frac{L_{\ast}}{L_{\sun}}
\frac{M_{\sun}}{M_{\ast}}\frac{\rho_g}{\rm g\ cm^{-3}}$} depends on
the stellar fundamental parameters (mass and luminosity) and grain
properties (size and density).  Using astronomical silicates
\citep{laor93} with various densities ranging from 1.8--3.3
g~cm$^{-3}$ to account for porosity, the blowout size (radius) is
$\sim$2.5--5 \um around a 2.25 $M_{\sun}$ star with a luminosity of
17.4 $L_{\sun}$. The maximum grain size is set at 1 mm since grains
larger than this size contribute negligible emission in the
infrared. We further assume the warm dust belt is confined in a narrow
ring (NR) with a fixed width of 2 AU. The exact function of the
density distribution in the warm belt has little impact in the output
SED for such a narrow ring; therefore, we adopt a Gaussian profile for
the ring's density distribution. Because we lack good constraints on
the Rayleigh-Jeans side of the excess emission, the SED models are not
well constrained even with all the assumptions mentioned before. A
small range of $\sim$8--15 AU with various combinations of minimum
grain size and size distribution slope can all provide satisfactory
fits to the observed SED. We illustrate this degeneracy by presenting
two narrow-ring models peaked at 9 and 13 AU shown in Figure
\ref{fom_innerexcess_models}, where both models have the same grain
size distribution (5 \um to 1 mm silicate grains with a size slope of
$-$3.65).

\begin{figure*} 
  \figurenum{4}
  \label{fom_alma_sim} 
 \epsscale{1.15} 
  \plotone{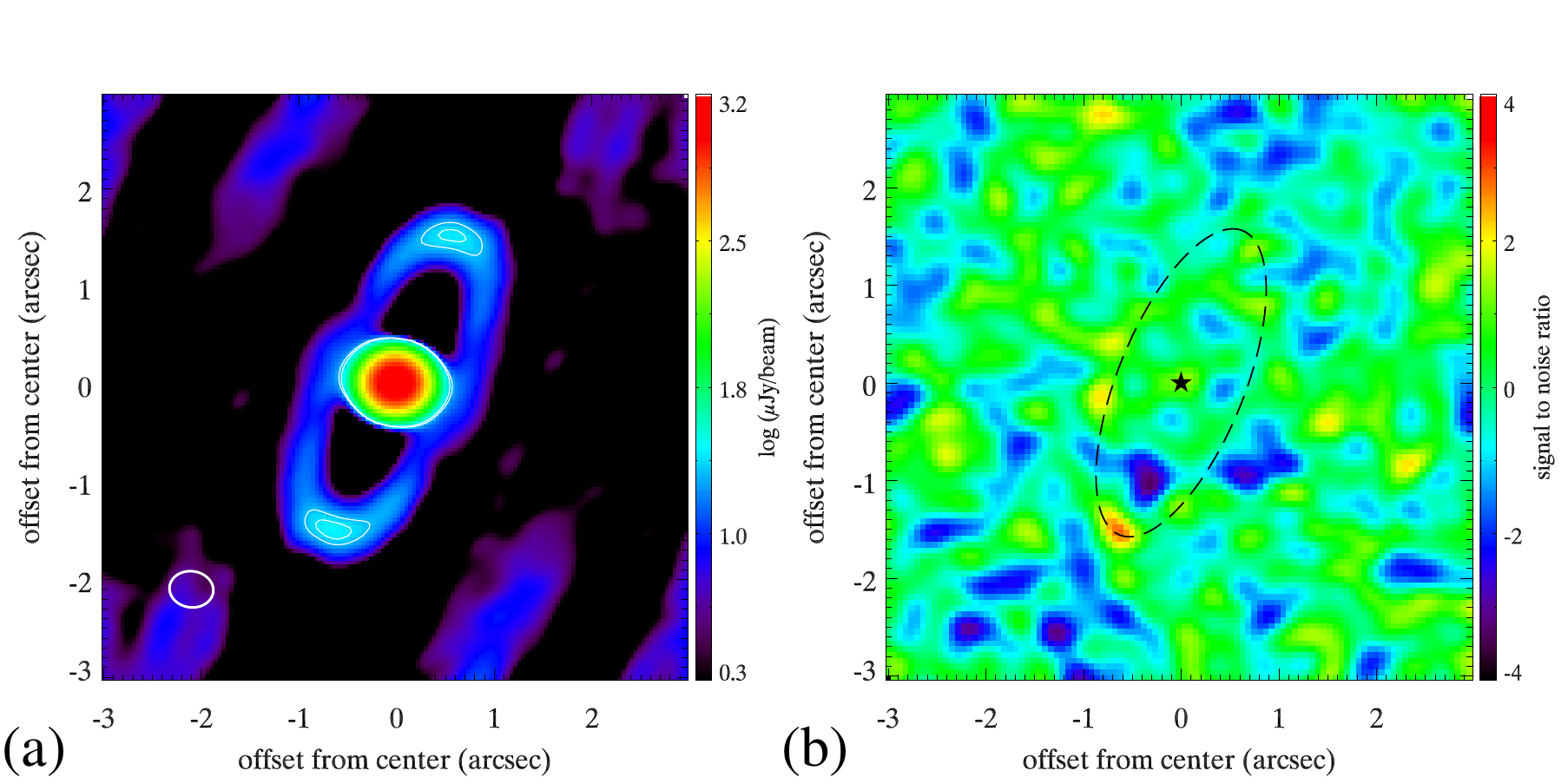}
  \caption{(a) Synthesized image of our model for a 13-AU narrow belt
at the same depth as the cycle 1 observation (details see Section 4.2)
where the synthesized beam is shown as the white ellipse at the
left-hand corner. No random thermal noise is included in this
simulation. The contours mark the inner belt at the levels of 20 and
36 $\mu$Jy beam$^{-1}$. (b) Synthesized inner belt image with star
subtraction. Random thermal noise is added to the simulated image (a)
first, then the same procedure that produced the Figure
\ref{fom_alma_cycle1}(b) (a point source fitting and subtraction in
the $uv$ plane) was applied to generate this model residual image. The
south-east disk ansa appears to be at $\sim$2$\sigma$ levels based on
this simulated model.  }
\end{figure*}

Assuming the inner belt is inclined by 66\arcdeg\ from face-on like
the outer cold belt, a narrow ring of 9 (13) AU would cover $\sim$13
(17) beams in its circumference in our cycle 1 map. Given a rms of 26
$\mu$Jy~beam$^{-1}$, the corresponding 3$\sigma$ upper limit for the
whole inner belt is 1 mJy and 1.3 mJy at 870 \um for such rings. Our
narrow-ring SED models give a much fainter total flux (see Figure
\ref{fom_innerexcess_models}), consistent with no detection. As
presented in Section 3.1, there are a few 1--2$\sigma$ blobs along the
expected 13-AU-ring circumference. We simulate the 870 \um model image
of the 13-AU narrow belt under the same observing depth and array
configuration as the cycle 1 data. Figure \ref{fom_alma_sim}(a) shows
the {synthesized} image without including random thermal noise. We
then added the random thermal noise by adjusting the PWV parameters in
$simobserve$ until the measured noise level reaches 26 $\mu$Jy
beam$^{-1}$ as in our cycle 1 observation. We then applied the same
procedure that generated Figure \ref{fom_alma_cycle1}(b) to generate a
star-subtracted residual map of the synthesized image shown in Figure
\ref{fom_alma_sim}(b) for comparison.  Note that the noise in our
simulation is added randomly; therefore, it is not expected to
coincide with the observation when compared to Figure
\ref{fom_alma_cycle1}b. Futhermore, we did not include any
background confusion as estimated by \citet{oteo15} using ALMA
calibration data, i.e., Figure \ref{fom_alma_cycle1}b is expected to
have some contribution of faint ($\gtrsim$3) background galaxies. 
Our simulation suggests that the south-east disk ansa might be
detected at $\sim$2$\sigma$ levels. Nevertheless, a deeper map is
needed to confirm this putative 13-AU belt.

\subsection{Complexity of Inner Debris Distribution -- A New Interpretation of Interferometric Data}

\subsubsection{A New Model for the Inner 20 AU Excess in Fomalhaut} 

Our proposed asteroid belt (a narrow belt near the water ice line,
presented in Section 4.2) would not produce any detectable signal in
the KIN measurements because the dust in this narrow belt is too cold
(i.e., too faint) at 8--13 $\mu$m, and it is located mostly outside
the field of view of KIN. To reconcile with the KIN measurements, here
we explore a different scenario including a P-R drag-in component from
the asteroid belt and a hot ($\sim$1500 K) ring produced by the
magnetically trapped nano grains proposed by Rieke et al.\ (2015,
accepted). To roughly estimate the feasibility of a P-R component from
the proposed asteroid belt, we first compare the typical P-R timescale
with the collisional one in our proposed ice-line belt\footnote{The
collisional time scale is formulated as $t_{\rm coll}=\frac{\sqrt{r^3
G M_{\ast}}}{2 \tau_{\rm eff}}$ and the P-R time scale is as $t_{\rm
PR}=\frac{c r^2}{4 G M_{\ast} \beta}$ where $r$ is the stellocentric
distance, $\tau_{\rm eff}$ is the optical depth.}. The collisional
time scale is $\sim$2$\times10^4$ yr for a belt at 10 AU with an
optical depth of 1$\times10^{-4}$ (see below) around a 2.25 $M_{\sun}$
star. This collisional timescale is slightly shorter than the P-R
timescale, $\sim$3.6$\times10^4$ yr for $\beta=$0.5 grains, suggesting
some amount of material can drift inward under the influence of P-R
drag, forming an interior extended disk. Additional material might be
deposited in this region by disintegrating comets; hence the P-R drag
component represents a rough lower limit but one that can be analyzed
without introducing uncontrolled free parameters.

We construct individual SED models for each of the
components according to its spatial constraint and expected grain
population in the drag-in disk. By adjusting the individual
contribution of each component, we then simultaneously obtain good
fits in the KIN measurements and the overall SED.

The amount of material that can be brought inward from a
dust-producing planetesimal belt due to P-R drag has been studied
analytically \citep{wyatt05} and numerically \citep{vanlieshout14}.
Basically, it depends strongly on the amount of material (i.e., the
collision rate) in the planetesimal belt. Assuming a single grain size
in the belt and that the collisions are destructive, the effective
optical depth ($\tau_{\rm eff}(r)$, the vertical optical depth for a
face-on disk) is parametrized as equations (4) and (5) in Wyatt
(2005). To estimate $\tau_{\rm eff}(r_0)$ in Wyatt's
formulae\footnote{The analytical model by \citet{wyatt05} has very
simplified assumptions: grains with one single size and emitting like
blackbodies. It is then not straight forward to use the optical depth
from a SED model (a size distribution of grains with imperfect
absorption coefficient) in these analytical formulae.}, we use the
observed fractional luminosity ($f_d$) and the relationship,
$\tau_{\rm eff}(r_0) = \frac{2 f_d r_0}{\Delta r}$, given by
\citet{kennedy15} (where $r_0$ and $\Delta r$ are the belt's location
and width.). The $f_d$ values are $\sim$1.2$\times10^{-5}$ in our SED
models presented in Section 4.2; therefore, $\tau_{\rm eff}(r_0)$ is
in the range of $\sim$1$\times10^{-4}$.  For a planetesimal belt at
$\sim$10 AU around a 2.25 $M_{\sun}$ star with an initial optical
depth of 1$\times10^{-4}$, the value of $\eta_0$ (equation (5) in
Wyatt 2005) is $\sim$2, and equation (4), $\tau_{\rm
eff}(r)=1\times10^{-4}[1+8(1-\sqrt{r/10 AU}]^{-1}$, gives the maximum
amount of material that can spiral inward. \citet{vanlieshout14}
performed detailed numerical simulations by including a size
distribution of particles in a collision-dominated planetesimal belt,
and found that the amount of the material due to P-R drag is roughly a
factor of 7 lower than the simple analytical calculation. As suggested
by \citet{kennedy15}, one can simply scale the $\eta_0$ value by a
multiplicative factor $k$ where $k=$ 1/7 to match the numerical result
from \citet{vanlieshout14}.  In addition, \citet{vanlieshout14} also
found that the size slope in the drag-in component is expected to be
steep with a wavy distribution.  Following the wording suggested in
\citet{kennedy15}, we refer to Wyatt's analytical study as the low
collision case and van Lieshout's numerical result as the high
collision case (or collision-dominated). Furthermore,
\citet{kobayashi09} suggest a density enhancement (a pile-up effect)
can occur near the dust sublimation radius of a P-R drag-in disk
(i.e., the inner edge of the P-R disk). As a dust grain drifts close
to the sublimation radius and starts to sublimate (i.e., reducing its
size), the radiation force on the dust grain becomes stronger and
temporarily halts its inward migration; therefore, a ring near the
sublimation radius can form. The sublimation radius and the
enhancement factors depend on the grain composition. Refractory
materials like silicate and carbon grains can give an enhancement
factor up to $\sim$4--6 \citep{kobayashi11}.

Based on the theoretical models described above, we construct the SED
of the P-R component in two parts. The first part is the drag-in disk
component that has a constant surface density, starting from $\sim$10 AU
(the inner boundary of the asteroid belt) to $\sim$0.23 AU (the
sublimation radius for silicate-like grains when they reach $\sim$1300
K), and is composed of a population of astronomical silicates in a
power-law size distribution with a slope of $-$5.5, and with sizes
ranging from 3 to 10 $\mu$m.  The choices of grain size parameters
follow the recipe developed by \citet{wyatt05} and
\citet{vanlieshout14} where grains with $\beta=$ 0.5 are the most
dominant sizes as the product of collisional cascades, hence, they
contribute the most emission from the drag-in disk (see Figure 5 in
\citealt{vanlieshout14}). The second part is for the density
enhancement near the silicate sublimation radius. We place a narrow
ring at 0.23 AU with a width of 0.035 AU (i.e., $\Delta r/r\sim$0.15)
to mimic the pile-up effect. For simplicity, we adopt the same grain
parameters as in the drag-in disk for this pile-up ring.

\begin{figure} 
  \figurenum{5}
  \label{complex_model_optdepth} 
  \plotone{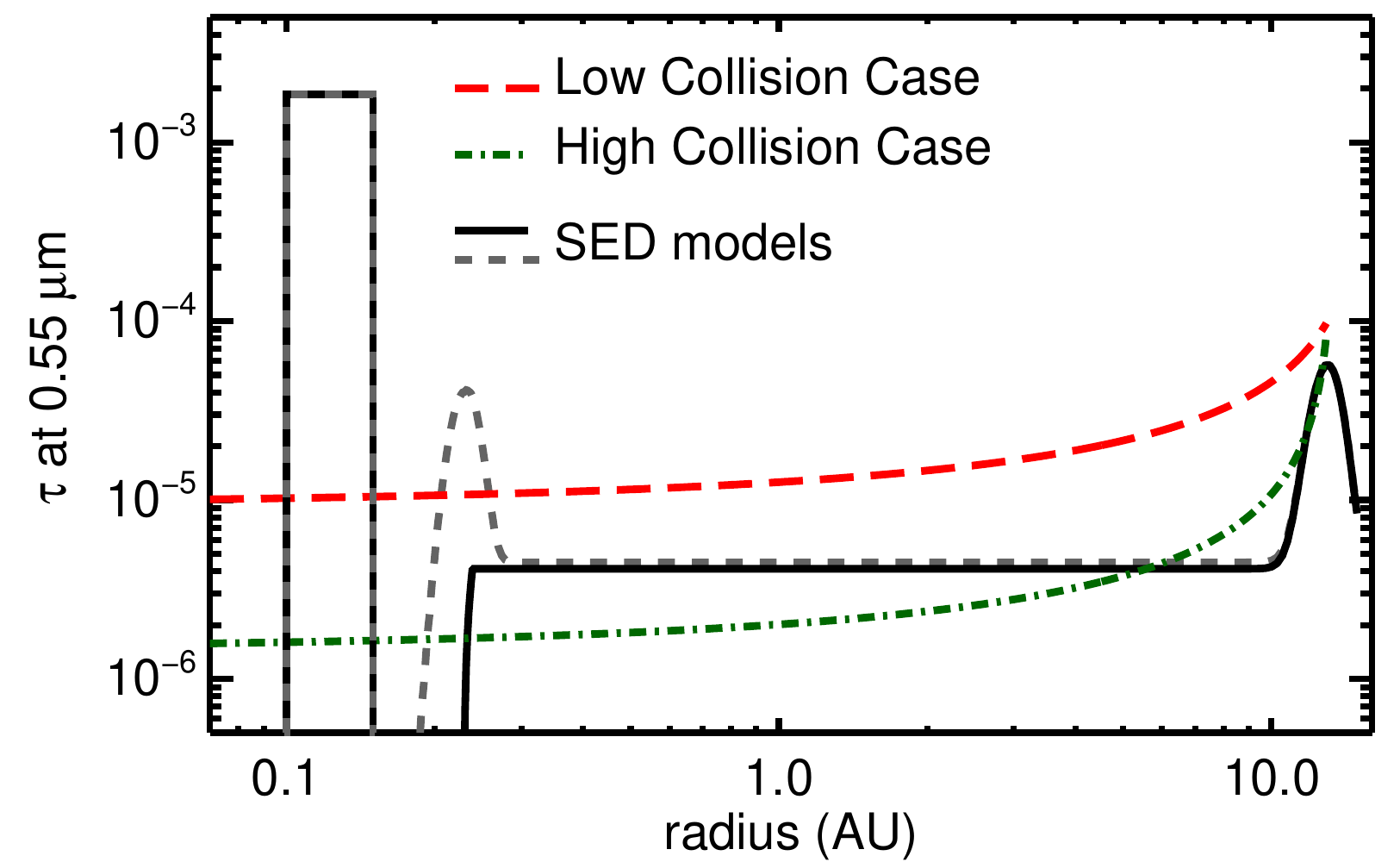}
  \caption{Optical depth in our SED models in comparison with
the low collisional rate case (long dash line) and high collisional
rate case (dash-dot line) calculations. The model with a pile-up ring
(upper panel of Figure \ref{complex_model}) is shown as the dash line,
and the model without a pile-up ring (bottom panel of Figure
\ref{complex_model}) is shown as the solid line. Both SED models have
the same asteroid belt (at $\sim$13 AU) and hot ring (at $\sim$0.1
AU).}
\end{figure}

We then adjust the amount of material in each of these components in
the combined SED: the asteroid belt (same as before, just different
normalization), the drag-in disk and the pile-up ring. To make sure
these SED parameters produce a vertical optical depth distribution
that is consistent with the theoretical expectation, we also compute
the corresponding optical depth in the SED model as following:
\begin{equation} \tau(r) = \frac{\sigma n dr}{2 \pi r} = \Sigma(r)
\int\limits^{a_{max}}_{a_{min}} Q_{abs} \pi a^2 f(a) da,
\end{equation} where $n$ is the dust  number density and $\Sigma(r)$ is the
corresponding surface density distribution (i.e., $\Sigma(r) = \int n dz$
), $\sigma$ is the cross section of a particle, i.e., $2 \pi a^2
Q_{abs}$ for a grain radius of $a$ and absorption coefficient
$Q_{abs}$, and $f(a)$ is the normalized grain size distribution. The
integration limits are the grain size boundaries used in the SED model;
i.e., for the drag-in disk, the limits are 3 and 10 $\mu$m for
astronomical silicates.  We assume a constant surface density
in the P-R disk; therefore, the resultant optical depth is flat for this
part. Figure \ref{complex_model_optdepth} shows the model optical
depth distribution for  two of the fits in comparison with the
theoretical values for the low and high collisional rate cases. As
shown in Figure \ref{complex_model_optdepth}, the material required in
the SED fit for the P-R disk is lower than the maximum amount from the
low collision case, but higher than the collision-dominated case at
distances far way from the asteroid belt.

The final component in our model is the narrow ring that gives rise to
the $K$-band excess. According to the modeling in 
\citet{rieke15}, nano grains can be photo-electrically charged and
magnetically trapped inside the dust sublimation radius ($\sim$0.23 AU
for silicates or $\sim$0.1 AU for carbon-like grains around Fomalhaut)
under a nominal condition (a dipole magnetic field of 1 G around an
A-type star), and the gyroradius (i.e., the inner edge of the trapped
particles) could get as close as 0.05 AU. For simplicity, we adopt
a constant-surface-density ring from 0.1 to 0.15 AU, composed of
amorphous carbon nano grains with sizes ranging from 0.01 \um (10 nm)
to 0.05 \um (50 nm) in an $a^{-4}$ size distribution. The exact grain
composition and size parameters do not have a huge impact on the SED
model as long as they produce a Rayleigh-Jeans-like spectrum from
$\sim$2--10 \um that best fits the $K$-band interferometric
measurement.

\begin{figure} \figurenum{6}
  \label{complex_model}
\plotone{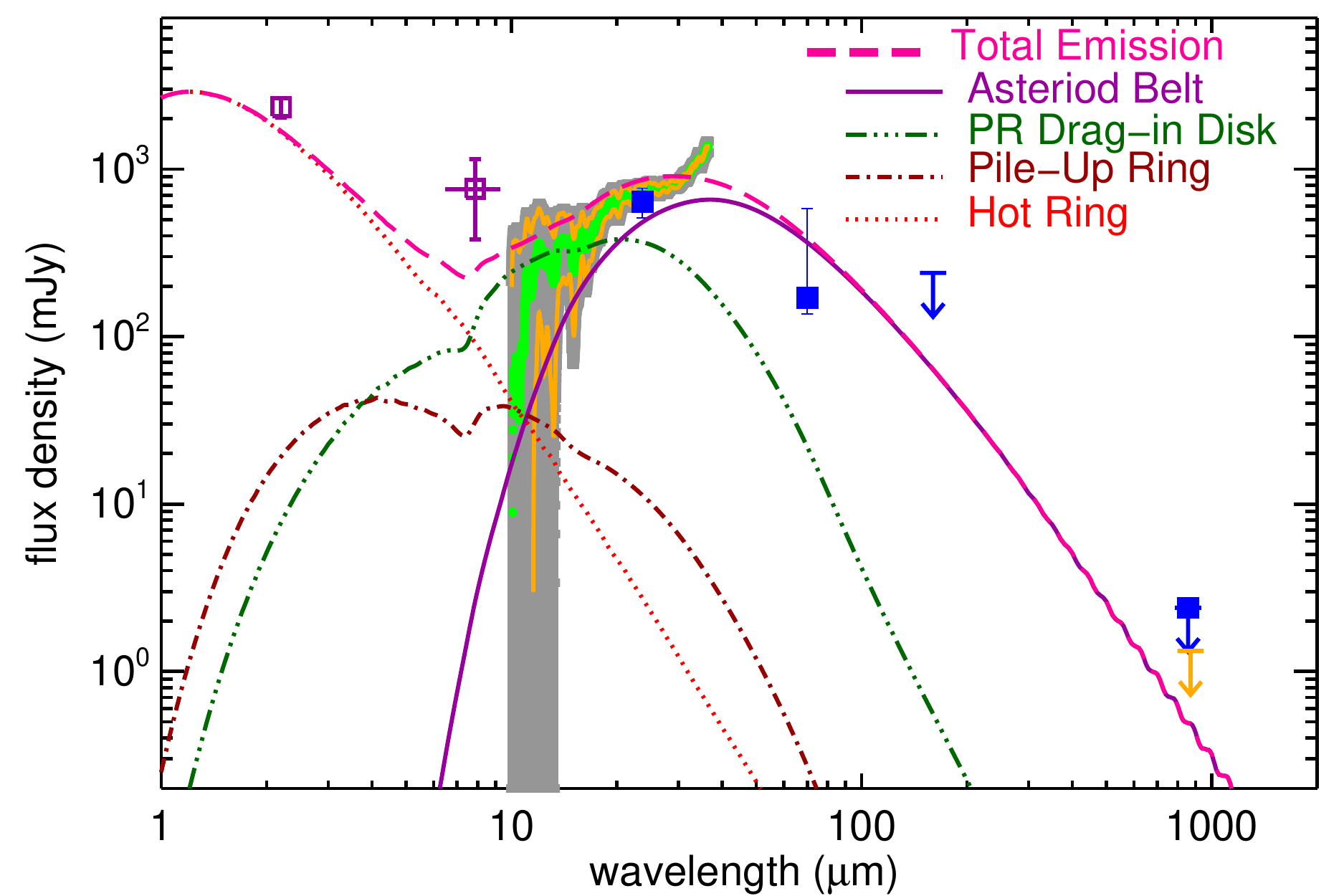}
\plotone{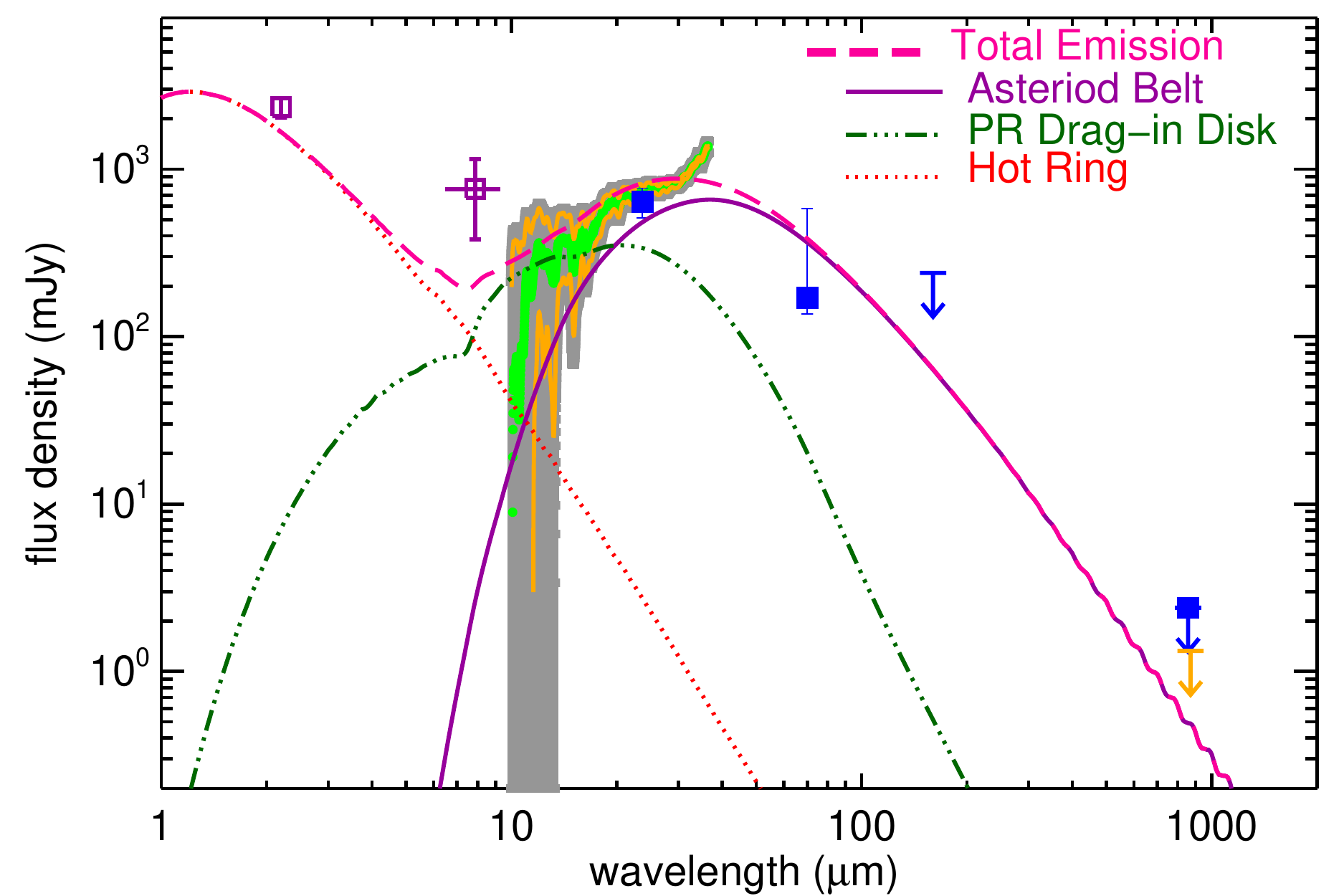}
  \caption{SED models to fit all the excess measurements in
Fomalhaut's inner 20 AU region. The data points are the same as in
Figure \ref{fom_innerexcess_models}. The total emission (thick dashed
line) is the sum of four different components: a 13-AU narrow asteroid
belt (solid line), a flat disk due to P-R drag (triple-dot dashed
line), and a hot ring composed of magnetically trapped nano grains
(dotted line). The top panel shows the model that includes a narrow
ring due to the pile-up effect near the inner edge of the P-R disk
(dot-dashed line) and the bottom panel shows the same model but
without the pile-up ring.}
\end{figure}

\begin{figure} 
  \figurenum{7}
  \label{KIN_comp} 
  \plotone{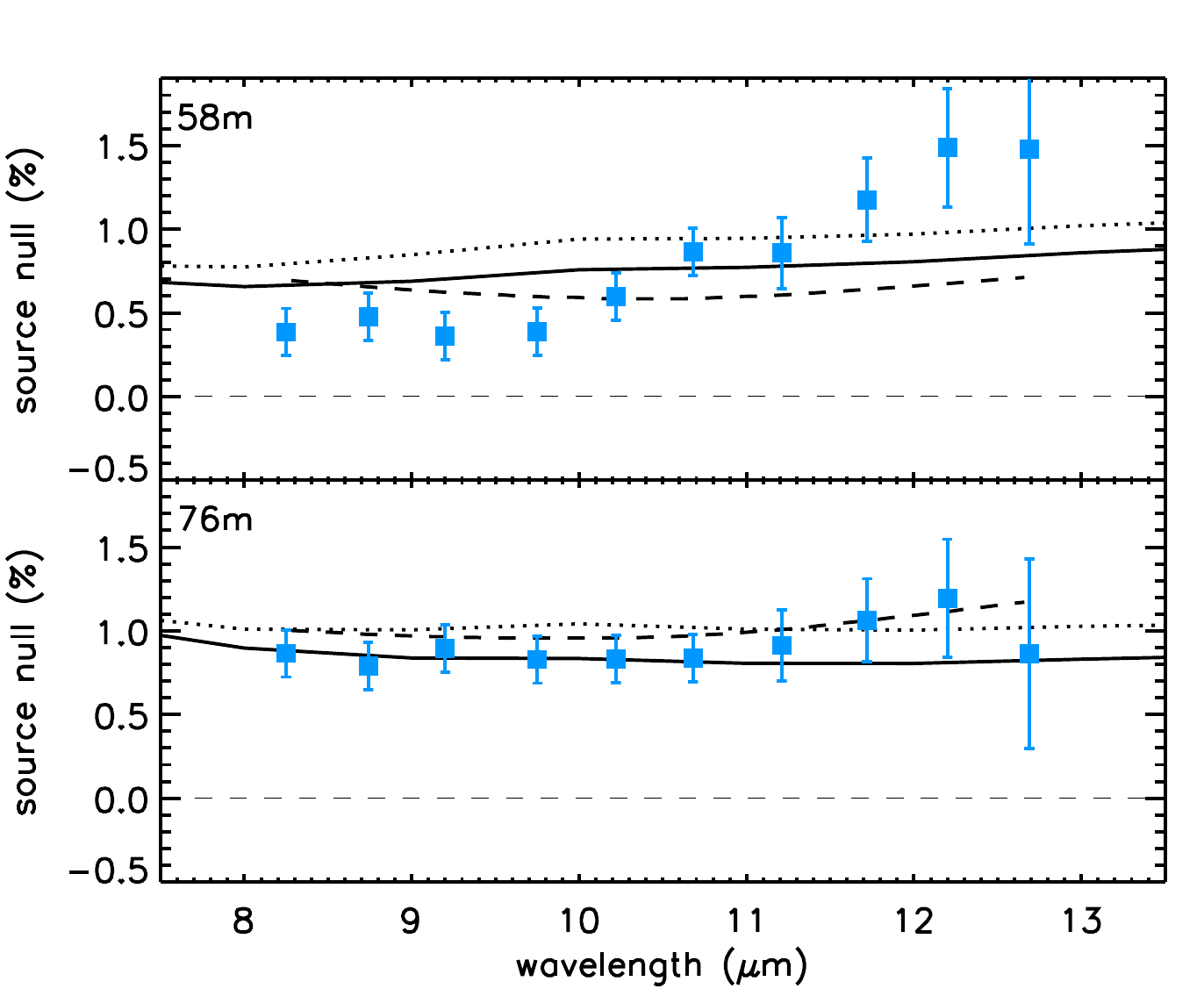}
  \caption{Null levels measured by the KIN in 2008 (squares with
error bars) from \citet{mennesson13} in comparison with the model null
levels (lines). The upper panel shows the short (58 m) baseline
result, while the lower one is for the long (76 m) baseline result. Our
best-fit models are shown as the solid line (without a pile-up ring,
i.e., the bottom panel of Figure \ref{complex_model}) and the dotted
line (with a pile-up ring, i.e., the top panel of Figure
\ref{complex_model}). The null levels from the \citet{lebreton13}
model are also shown (the thick dashed line) for comparison.}
\end{figure}

There are four individual components in our final SED model: the
asteroid belt, the P-R disk, the pile-up ring, and the magnetically
trapped hot ring. As a result, many combinations of the scaling for
each of the components (i.e., the total dust mass) give satisfactory
fits in the overall SED, and Figure \ref{complex_model} shows two of
them.  Our final constraint comes from the spatial information in the
KIN measurements. Each of the good-fit SED models produces a slightly
different spatial flux distribution at the KIN wavelengths (8--13
$\mu$m), i.e., different wavelength-dependent null levels (the
fraction of transmitted flux detected by KIN) at different baselines
(resolutions). We first simulate the high-resolution, face-on model
images at 7--14 \um using the best-fit SED models. Assuming that the
inner dust structures have the same inclination and position angles as
the cold belt ($i$=66\arcdeg\ and P.A.=156\arcdeg), the model images
are then inclined and rotated accordingly. Using these model images,
we then compute the expected null levels on the given dates and
baselines of the KIN observations (Table 2 of \citealt{mennesson13}),
and compare them with the observed levels. Figure \ref{KIN_comp} shows
examples of the comparison.

By changing the combination of individual components and the
associated SED parameters (like a Gaussian ring vs.\ a flat disk), we
explore the parameter space that the KIN measurements are sensitive
to. We find that (1) an extended dust component in the $\sim$AU region
(the proposed P-R disk) is needed to explain the KIN measurements,
i.e., a hot nano-grain ring at $\sim$0.1 AU is not enough to produce
the KIN signals, corroborating the finding in \citet{mennesson13}. (2)
The KIN measurements are not sensitive to the changes in the dust
structure within $\sim$0.3 AU (i.e., the hot and pile-up rings);
especially, the model with a pile-up ring simply gives slightly higher
null levels than those of the model without.  Our SED model for the
pile-up ring (the upper panel of Figure \ref{complex_model})
represents the maximum amount of emission in the model that is
consistent with the overall SED. This model gives an enhancement
factor of $\sim$10 in the optical depth, which is the most extreme
case found in the study of \citet{kobayashi11} while a typical
enhancement around an A-type star for silicate grains is $\sim$3.
Since the expected emission from the maximum pile-up ring is very
minimal compared to the nano-grain hot ring and the P-R disk (see the
upper panel of Figure \ref{complex_model}), the available measurements
cannot determine whether such a pile-up ring exists. (3) The
wavelength-dependent behavior in the KIN short baseline measurements
(the upper panel of Figure \ref{KIN_comp}) is difficult to match with
our simple axi-symmetric model. Future observations from LBTI
\citep{hinz12} can help to confirm this wavelength-dependent behavior,
and provide complementary spatial constraints that are crucial to
distinguish between degenerate KIN models (e.g., \citealt{defrere15}).

Table \ref{tbl:sedparameters} summarizes the SED parameters in each of
the components. Note that the current data do not put a strong
constraint on the existence of a pile-up ring near the silicate
sublimation radius; we list its SED parameters just for the sake of
completeness. With or without the ring due to pile-up, our model
appears to give a satisfactory fit to the KIN data  (Figure \ref{KIN_comp}), 
as does also the Lebreton model (details see Section 5.3).

\begin{deluxetable*}{llccclcclc}
\tablecolumns{10}
%\tablewidth{0pc}
\tablecaption{Parameters in the four SED components \label{tbl:sedparameters}}
\tablehead{ 
\colhead{Component } & \colhead{Dust Type} & $a_{\rm min}$ & $a_{\rm max}$ & $q$  & Density Type$^{\dagger}$ & $R_{\rm in}/R_{\rm p}$  & $R_{\rm out}/R_{\rm w}$ & Dust Mass & $f_d $  \\
\colhead{}  & \colhead{} & [$\mu$m] & [$\mu$m]& \colhead{} & & \colhead{[AU]} & \colhead{[AU]} & \colhead{[$M_{\earth}$]}  &  
}
\startdata 
Asteroid Belt & Silicates & 5 & 1000 & $-$3.65 & Gaussian Ring & 13 & 2 & 1.5$\times10^{-5}$ & 8.8$\times10^{-6}$ \\ 
P-R Disk       & Silicates & 3 & 10   & $-$5.5 & Flat Disk & 0.23 & 10  & 1.4$\times10^{-7}$ & 1.3$\times10^{-6}$ \\ 
Pile-up Ring$^{\ddagger}$  & Silicates & 3 & 10   & $-$5.5 & Gaussian Ring & 0.23 & 0.035 & 7.1$\times10^{-10}$ & 5.2$\times10^{-6}$ \\ 
Hot Ring  & Carbon & 0.01 & 0.05  & $-$4.0 & Flat Disk & 0.1 & 0.15 & 8.0$\times10^{-11}$ & 9.7$\times10^{-4}$
\enddata 
\tablenotetext{$^\dagger$}{We use two types of surface density
distribution: (1) a Gaussian ring characterized as the peak radius
($R_{\rm p}$) and the width ($R_{\rm w}$), and (2) a flat disk
characterized as the inner ($R_{\rm in}$) and outer ($R_{\rm out}$)
radii.}
\tablenotetext{$^{\ddagger}$}{Current data do not put a
strong constraint on the existence of a pile-up ring.} 
\end{deluxetable*}

\section{Discussion}

\subsection{Does the Derived Dust Mass Make Sense?}

Is the amount of dust in the asteroid belt too much for a belt 
undergoing normal collisional evolution? A simple way to answer this
question is to compare the observed fractional luminosity with the
maximum fractional luminosity expected from collisional evolution
models, $f_{\rm max}$ (i.e., equation (21) from
\citet{wyatt07}). After scaling for the mass and luminosity of
Fomalhaut, we find $f_{\rm max}$ is $\sim$1$\times10^{-5}$ for a belt
at 10 AU at an age of 450 Myr. The observed fractional luminosity is
$\sim$0.8$\times10^{-5}$ derived from our model, implying the amount
of dust observed in the belt does not violate the collisional
evolution models\footnote{The criterion is usually set when $f_d>$100
$f_{\rm max}$ \citep{wyatt07}.}. The $\sim$10 AU ice-line belt can be
an in-situ planetesimal belt.

The optical depth distribution in our simple P-R disk model (shown in
Figure \ref{complex_model_optdepth}) is between the low and high
collision cases expected from theoretical calculations except for the
region right interior of the inner belt. We also used a similar
gradual curve in that part of the P-R disk, but it made no difference
in the output SED and resultant KIN nulls. This is not a surprise
given that this region is outside the field of view of KIN, and SED
models are not sensitive to changes in small spatial scale. The total
derived dust mass in the P-R disk is 1.4$\times10^{-7} M_{\earth}$,
which is $\sim$100 times less than the derived dust mass in the
asteroid belt (1.5$\times10^{-7} M_{\earth}$), suggesting the ice-line
belt can sustain the mass needed in the P-R disk. Overall
the derived dust masses in the asteroid belt and the P-R disk agree
with the collisional and dynamical evolution.

For the hot ring, our derived mass is 8$\times10^{-11}
M_{\earth}$. This value is very similar to the mass ($\sim$$10^{-10}
M_{\earth}$) derived from other studies
\citep{absil09,lebreton13,rieke15}.  In other words, the mass required
to explain the VLTI $K$-band excess is on the order of $\sim$$10^{-10}
M_{\earth}$. The question remains whether the P-R disk can supply
enough material to form a nano-grain hot ring. We can estimate the
mass flow near the inner boundary of the P-R disk (without a pile-up
ring) as $\dot{M}_{\rm PR}(r)=\frac{L_{\ast}}{c^2}Q_{\rm pr} \tau(r)$
in the low collisional rate case (equation (9) from
\citet{vanlieshout14}). With $L_{\ast}=$ 17.4 $L_{\sun}$, $Q_{\rm
pr}=$ 1 for $\sim$$\mu$m-size grains, and $\tau\sim$ 4$\times10^{-6}$
from our model, the mass flow rate is
$\sim$2$\times10^{-12}M_{\earth}$yr$^{-1}$ at 0.23 AU (where P-R
grains sublimate). In comparison, the maximum mass flow rate from P-R
drag at 0.23 AU is 3$\times10^{-11}M_{\earth}$yr$^{-1}$, using van
Lieshout's equation (11). These rates and the mass required in the hot
ring suggest a typical resident time of $\sim$5--50 years for the nano
grains. The lifetime of the trapped nano grains is estimated to be
$\sim$months to years depending on the grain sizes
\citep{rieke15}. The estimated resident time is longer than the
lifetime of nano grains if they are only supplied from the P-R
disk. This implies that the P-R disk in our model can be a source to
supply the nano grains; however, other mechanisms like star-grazing
comets might be needed to supplement this mechanism.

\subsection{How Sensitive are These Models to the Adopted Grain Compositions?}

The choice of the grain composition in the hot ring does not really
affect the outcome of the model as long as the output SED is similar
to Rayleigh-Jeans or steeper (details see \citet{rieke15}). 
We adopt amorphous carbon grains because they are a
common material that has a high sublimation temperature. Other oxides
like FeO can also work as well (see the discussion in \citealt{su13}
and \citealt{rieke15}).

We adopt only one composition, the silicate-like grains (astronomical
silicates), in both the asteroid belt and the P-R disk because they
are the most common material found in interplanetary dust particles
(IDPs). As a result, the model SEDs show weak bumps in the 10 and 20
\um silicate features because the smallest size in the model is
$\sim$3 $\mu$m, just small enough to give rise to such features. A
mixture with modest amounts of organic material would also produce a
similar result with additional adjustments (like the dust sublimation
radius) on the exact grain properties. The weak silicate features will
be washed out when using a mixture with other featureless
material. The exact numbers of the best-fit grain properties and
density distribution are expected to vary, depending on the adopted
grain compositions. The overall behavior in the link between the
dust-producing asteroid belt and the interior, low-density, P-R disk
remains the same.

\subsection{Comparison Between Our and Lebreton's Models}

Lebreton et al. (2013) have presented the most complete and detailed
model of the Fomalhaut inner debris disk to date. Since their model is
different from ours, it is important to clarify what data are included
in their fit and the main difference between the two models.  The
primary data used to constrain Lebreton model come from the spatial
constraints from the KIN 8--13 $\mu$m measurements, the VLTI K-band
excess, and the broadband photometry at 24 and 70 $\mu$m. They propose
two distinct dust populations in the Fomalhaut inner disk: (1) a very
narrow ring confined at 0.1--0.3 AU from the star and composed of
unbound, very small 0.01--0.5 $\mu$m carbon-like grains, and (2) a
disk peaked at $\sim$2 AU with a $r^{-1}$ density distribution outward
and composed of silicates and carbon grains in a very steep power-law
size distribution (index of $-4.8$ to $-4.1$) with a minimum cutoff of
$\sim$ 2--3 $\mu$m. Figure 3 also shows their model SED. This model
clearly over-predicts the flux in the 15--30 $\mu$m range (the IRS
data were not used as a constraint), and the 24 $\mu$m excess is
$\sim$1 Jy (see their Figure 5), which is higher than the unresolved
excess of 0.64 $\pm$0.13 Jy estimated in Section 4.1. This is likely
due to their lower photospheric estimate, $\sim$2.78 Jy at 24 $\mu$m
(see their Figure 5).

Several scenarios were discussed and explored in Lebreton et
al. (2013) to explain the sources of dust in the $\sim$2-AU region,
which include: 1.) in-situ dust production, i.e., a planetesimal belt
at 2 AU\footnote{The typical dust temperatures in this region are
$\sim$400--450 K, which is very different from the dust temperature we
refer as the ice-line belt.}; 2.) P-R drag-in grains from the cold
($\sim$50 K) outer belt; and 3.) planetesimal delivery scattered
inward by multiple planets inside the cold belt. They ruled out the
in-situ dust production since the observed fractional luminosity is a
few orders of magnitude higher than what a 2-AU belt in collisional
equilibrium can produce at the age of Fomalhaut (450 Myr). In
comparison, the derived dust properties in our ice-line belt at
$\sim$10 AU region are consistent with collisional evolution of an
in-situ planetesimal belt as discussion in Section 5.1.  The P-R drag
from the outer belt was also ruled out because the amount of drag-in
grains is inadequate to explain the dust level at the 2 AU
region. They conclude that inward scattering of planetesimals from the
outer belt by a chain of tightly packed planets is marginally
adequate.

The innermost, hot ring poses more challenges as in other previous
studies because its SED requires emission dominated by very small
grains that are unstable against photon pressure blowout.
\citet{lebreton13} include a detailed treatment for the dust
sublimation process to produce the very small grains in the hot ring
and derive a replenishing rate of
$\sim$8$\times10^{-8}M_{\earth}$yr$^{-1}$ to sustain the hot ring.
Conventional mechanisms to supply such a flow fall far short of this
value; for example Lebreton et al.\ find that, P-R drag from the 2 AU
belt is inadequate by nearly four orders of magnitude. Even invoking
an evaporating planet came up short, although the yield from such an
event is very uncertain. \citet{bonsor14} developed the comet
scattering hypothesis further and found that, with some very
closely-packed inner planets and an outermost planet migrating into
the cold planetesimal belt, it could supply a rate slightly more than
$10^{-9}M_{\earth}$yr$^{-1}$, still an order of magnitude too small.
In addition, this hypothesis depends on an ad hoc arrangement of
planets and is implausible to be operating equally well around the
many other stars with similar very hot excess emission.

To mitigate these problems, \citet{lebreton13} proposed that the very
small grains are trapped against escape by gas in the hot ring, but
this solution has significant problems. Such a gas disk with normal
(hydrogen-rich) abundances would be easily detected in optical
emission lines (it would be analogous to the disks around Ae/Be stars
-- see, e.g., \citealt{mendigutia15}). Instead, \citet{lebreton13}
suggest that the gas is the residue from the sublimation of dust, and
show that a total mass of 5$\times 10^{-3} M_{\earth}$ would then
suffice. In this case, emission in the C\ion{II} 158 $\mu$m line would
be expected \citep{zagorovsky10}. \citet{cataldi15} have used {\it
Herschel}/PACS to place an upper limit on the emission from this line
from the entire Fomalhaut ring system, and show that the signal from
the spaxal centered on the star (see their Figure 1) is even less than
from the rest of the cold ring. Assuming the gas coincident with the
very hot dust would be at a similar temperature, the upper limit is an
order of magnitude lower than the required mass in the gas disk
hypothesized by \citet{lebreton13}.

In our model, the excess detected by the long-baseline KIN observation
arises from the P-R drag-in grains from the $\sim$10-AU ice-line
belt. These drag-in grains then sublimate as they get close to the
sublimation radius (0.23 AU for silicate-like grains) and become the
charged nano grains trapped by the the weak magnetic field of the star
as proposed by \citet{rieke15}, forming the hot ring. The lifetime of
these nano grains (months to years) then suggests a replenishing rate
of $\sim$$10^{-10}M_{\earth}$yr$^{-1}$.  As discussed in Section 5.1,
the P-R mass inflow rates are in the range of $10^{-12}-10^{-11}
M_{\earth}$yr$^{-1}$ from the ice-line belt; an additional source of
nano-grains, like star-grazing comets, may be needed to supplement the
P-R inflow from the ice-line belt. This star-grazing comet delivery is
different from the cometary delivery scenario presented in
\citet{bonsor14} where a very specific planetary configuration is
required to sustain the high influx rate of comets. 
There will be a wide range of planetary
configurations that both meet the scattering criteria as evaluated by
\citet{bonsor12} and allow to scatter a few star-grazing comets
per year into the inner part of the Fomalhaut planetary system
(including the tightly pack \citet{bonsor14} configuration).

\subsection{The Ice-line Asteroid Belt -- A Natural Source for the Dust Interior to the Belt }

The prominent cold belt cannot supply enough material to the inner
2-AU region as discussed by \citet{lebreton13}. With our new
parameters for the ice-line belt (dust mass of 1.5$\times10^{-5}
M_{\earth}$ and collisional time scale of 2$\times10^{4}$ yr), the
required mass flow rate, $\sim$$10^{-9}M_{\earth}$yr$^{-1}$, is still
too high to be sustained by the outer 140 AU cold belt due to PR drag
(the maximum mass flow is $\sim$$10^{-12}M_{\earth}$yr$^{-1}$,
equation (11) from \citealt{vanlieshout14}). Unlike $\epsilon$ Eri
\citep{macgregor15}, the 7 mm ATCA observation of Fomalhaut
\citep{ricci12} does not reveal any convincing excess emission above
the photosphere.  Also ionized winds from A-type stars like Fomalhaut
are extremely weak \citep{babel95}.  Therefore, it is unlikely that
stellar-wind drag can aid the P-R drag significantly. An in-situ
ice-line belt, as we have demonstrated, is the best alternative to
supply the interior dust detected by infrared interferometry.

The amount of interior dust that can be delivered from a planetesimal
belt may even be overstated by the P-R drag estimates.  In our Solar
System, the dust level interior to the Kuiper belt is thought to be
constant into $\sim$10 AU, where most of the particles are ejected by
Saturn and Jupiter \citep{moro-martin05}, and only a very small amount
of the P-R drag-in dust can reach Earth's vicinity
\citep{liou99,kuchner10,vitense10}. These model predictions are
consistent with the measurement by the Student Dust Counter
\citep{szalay13,szalay15} on board the New Horizons Mission. In
addition to the denseness of a belt (collision rate), giant planets
located interior to a planetesimal belt can also reduce the amount of
dust interior to the belt. To verify whether the amount of material
due to P-R drag interior to a planetesimal belt is lower than the
expected value (i.e., a sign of an additional depletion mechanism like
the presence of giant planet(s)), a detailed collisional cascade
calculation is needed. However, such a model requires detailed
information about the planetesimal belt (i.e., location and mass) that
is currently lacking.

\section{Conclusion}

We report an ALMA cycle 1, 870 \um observation centered at the star
position of the Fomalhaut planetary system. We detect a point source,
consistent with the bare stellar photosphere, and no extended structures
that are brighter than 3$\sigma$ in the central 15$\times$15 AU
region. We evaluate all available measurements to constrain the
infrared excess arising from dust in the inner 20 AU region in
Fomalhaut, and conclude that a dust-producing planetesimal belt at the
ice sublimation temperature (i.e., an asteroid-belt analog) is the
most likely origin for the infrared excess longward at $\sim$15
$\mu$m. The location of the ice-line belt is estimated to be at
$\sim$8--15 AU using SED models with nominal parameters for a narrow
belt with a 3$\sigma$ upper limit of total flux less than 1.3 mJy at
870 $\mu$m. Assuming the inner belt has the same orientation as the
outer one (inclination and position angles of 66\arcdeg\ and
156\arcdeg), but centered at the star position, we detect a few
1--2$\sigma$ blobs along the expected disk circumference of a 13-AU
belt. Although our SED model suggests such a 13-AU belt might be detected
by our cycle 1 data at 1--2$\sigma$ levels, the putative 13-AU belt
needs future confirmation. 

We further propose a new coherent model to explain the interferometric
hot excesses by connecting the proposed asteroid belt to the dust
structures inside of it. We suggest that a small amount of material
from the ice-line belt can spiral inward under the influence of P-R
drag, forming an interior, extended disk composed of dust grains that
have $\beta$ values closer to 0.5. The inner boundary of the P-R disk
is set by the dust sublimation radius of the dominant material like
silicates ($\sim$0.23 AU around Fomalhaut when grains reach $\sim$1300
K). We also consider a possible pile-up ring near the silicate
sublimation radius as proposed by \citet{kobayashi09} with an
enhancement factor of a few in the optical depth. We show that the
required optical depth of such an interior disk (and a pile-up ring)
is lower than the maximum allowable amount through the P-R drag under
the theoretical calculations. Finally, the sublimation of these
drag-in silicate grains is likely to produce nano carbon-like or
FeO-like particles that should be blown out by radiation pressure very
quickly under a normal circumstance. However, a weak ($\sim$1 G)
magnetic field from a fast rotating A-type star like Fomalhaut can
efficiently trap these photoelectrically charged nano grains forming a
hot ring. The location of the hot ring depends on the magnetic field
strength and the grain sublimation temperature. As modeled by
\citet{rieke15}, a ring at $\sim$0.1 AU and composed of nano, very
refractory grains around Fomalhaut can explain the $K$-band excess
detected by VLTI. The required resident time of the nano grains
sustained entirely from the P-R disk is somewhat longer than the
collisional lifetime of these nano grains; therefore, additional
sources like a few star-grazing comets are needed to explain the hot
ring.  Combining all these components (the asteroid belt, the P-R
disk with and without a pile-up ring, and the hot nano-grain ring), we
can simultaneously obtain good fits to the excess SED and satisfy the
spatial constraints set by the KIN measurements.

Studying the inner zones in debris disks requires high angular
resolution that can only be provided in scattered light by
ground-based high-contrast imaging facilities and/or {\it HST}, and by
ALMA in thermal emission. However, the scattered light study of the
inner debris structures around A-type stars like Fomalhaut is
challenging because the inner debris is intrinsically faint due to
lower mass (compared to the outer debris) and lack of efficient
scatterers of small grains (typical radiation blowout size for
Fomalhaut is $\sim$3 $\mu$m). Although future LBTI and {\it JWST}
mid-infrared observations can trace $\sim$$\mu$m-size grains in the
inner zone in great detail, their structures are much more influenced
by radiation forces (radiation pressure and P-R drag). Only ALMA can
resolve the detailed structure of the asteroid belt by tracing the
mm-size grains that show the imprint of extrasolar terrestrial planets
through dynamical interactions. Knowing the detailed properties of the
asteroid belt not only will help us better understand the processes
creating tenuous dust structures interior of the belt, but also
provides insights into extrasolar terrestrial planets that we cannot
probe otherwise.  Deeper observations with the full ALMA array have
the potential to reveal directly the asteroid belt inferred from our
SED models.

\acknowledgments

This paper made use of the following ALMA data: ADS/JAO.ALMA
2012.0.00238.S (PI: S.P.L.). ALMA is a partnership of ESO
(representing its member states), NSF (USA), and NINS (Japan),
together with NRC (Canada) and NSC and ASIAA (Taiwan), in cooperation
with the Republic of Chile. The Joint ALMA Observatory is operated by
ESO, AUI/NRAO, and NAOJ.  K.Y.L.S. is grateful for funding from NASA's
ADAP program (grant number NNX11AF73G).  Support for G.H.R. is
provided by NASA through contract 1255094 and 1256424 issued by
JPL/Caltech to the University of Arizona. S.P.L thanks the support of
the Ministry of Science and Technology (MoST) of Taiwan with Grants
NSC 98-2112-M-007-007-MY3, NSC 101-2119-M-007-004 and MoST
102-2119-M-007-004-MY3.

\end{document}